\documentclass[trackchanges,twocolumn]{aastex7}
\usepackage[utf8]{inputenc}
\usepackage{subfigure}
\usepackage{CJKutf8}

\usepackage{appendix}
\usepackage{amsmath} 
\usepackage{amsfonts} 
\usepackage{textcomp} 
\usepackage{csquotes}
\usepackage{soul}
\usepackage{xcolor}

\begin{document}

\title{\textbf{\Large Radio Follow-Up Observations of a Weak-Line Quasar Exhibiting Remarkable X-ray Variability}}

\author[orcid=0009-0002-5980-8437]{Ayushi Chhipa}
\affiliation{ Indian Institute of Astrophysics, Koramangala II block, Bangalore - 560034, Karnataka, India}
\affiliation{Pondicherry University, R.V. Nagar, Kalapet, 605014 Puducherry, India}
\email[show]{ayushi.chhipa@iiap.res.in}

\author[orcid=0000-0001-5937-331X]{M. Vivek} 
\affiliation{ Indian Institute of Astrophysics, Koramangala II block, Bangalore - 560034, Karnataka, India}
\email{vivek.m@iiap.res.in}  

\author[0000-0002-8070-5400]{Nayana A. J.}
\affiliation{Department of Astronomy, University of California, Berkeley, CA 94720-3411, USA}
\affiliation{Berkeley Center for Multi-messenger Research on Astrophysical Transients and Outreach (Multi-RAPTOR), University of California, Berkeley, CA 94720-3411, USA}
\affiliation{ Indian Institute of Astrophysics, Koramangala II block, Bangalore - 560034, Karnataka, India}
\email{nayana@berkeley.edu}

\author[orcid=0000-0003-3203-1613]{P. Kharb}
\affiliation{National Centre for Radio Astrophysics-Tata Institute of Fundamental Research (NCRA-TIFR), Pune-411007, India}
\email{kharb@ncra.tifr.res.in}

\author[orcid=0000-0002-0167-2453]{W. N. Brandt}
\affiliation{Department of Astronomy and Astrophysics, 525 Davey Lab, The Pennsylvania State University, University Park, PA 16802, USA}
\affiliation{Institute for Gravitation and the Cosmos, The Pennsylvania State University, University Park, PA 16802, USA}
\affiliation{Department of Physics, 104 Davey Laboratory, The Pennsylvania State University, University Park, PA 16802, USA}
\email{wnb3@psu.edu}

\author[orcid=0000-0002-5825-9635]{Preshanth Jagannathan}
\affiliation{National Radio Astronomy Observatory, 1003 Lopezville Road, Socorro, NM 87801, USA}
\email{pjaganna@nrao.edu}

\author[orcid=0000-0002-0367-812X]{Janhavi Baghel}
\affiliation{National Centre for Radio Astrophysics-Tata Institute of Fundamental Research (NCRA-TIFR), Pune-411007, India}
\email{jbaghel@ncra.tifr.res.in}

\author[orcid=0000-0003-1795-3281]{Savithri H. Ezhikode}
\affiliation{St. Francis de Sales College (Autonomous), Electronics City, Bengaluru - 560100, India}
\affiliation{Department of Physics and Electronics, CHRIST (Deemed to be University), Bangalore - 560029, India}
\email{savithrihezhikode@gmail.com}

\author[orcid=0000-0001-5356-1221]{C. H. Ishwara-Chandra}
\affiliation{National Centre for Radio Astrophysics-Tata Institute of Fundamental Research (NCRA-TIFR), Pune-411007, India}
\email{ishwar@ncra.tifr.res.in}

\begin{abstract}
SDSSJ1539+3954 ($z\approx 1.935$), a radio-quiet weak-line quasar (WLQ), exhibited exceptional X-ray variability in 2019--2020, with its X-ray flux increasing by over 20 times from 2013 to 2019 and subsequently dropping by at least a factor of nine in 2020. Motivated by the empirical correlations between X-ray and radio emission in AGN cores, we carried out a follow-up radio study in the 0.3--10 GHz range using GMRT (2020, 2022, 2024) and VLA (2022), and analyzed archival VLASS 3 GHz data (2017-2023) to investigate the source's radio properties and potential connection with the X-ray behavior. Our observations reveal a compact radio source with a spectral index of -0.65$\pm$0.15 in the frequency range of 0.3--1.4 GHz and -1.09$\pm$0.16 in 3--10 GHz. While the source was undetected in VLA-FIRST (1994) and VLASS epochs, the GMRT and VLA observations show no statistically significant variability over the monitored period. The absence of detectable changes in the radio flux, despite strong X-ray variability, suggests no direct connection between the X-ray variability and the radio emission, consistent with the Thick-Disk plus Outflow (TDO) model for WLQs. However, the sensitivity limit of the surveys prevents us from drawing definitive conclusions regarding longer timescale variability between the VLA-FIRST and GMRT epochs. We further explore possible mechanisms driving the radio emission from this source. Our analysis rules out small-scale jets and coronal emission as the primary drivers of the radio emission, suggesting that extended emission from AGN winds and star formation is the more plausible mechanism.
\end{abstract}

\section{Introduction}\label{sec:intro}
Active Galactic Nuclei (AGN) are classified into radio-loud and radio-quiet categories based on the \enquote{Radio--loudness ($R$)--parameter}, which is the ratio of flux densities observed at 6 cm and 4400 $\mathring{\mathrm{A}}$ wavelengths; an AGN is considered radio-loud if $R\ge $10 \citep{kellermann1994radio,kellerman1989}, otherwise, it is classified as radio-quiet. Nonetheless, more recent studies, including those by \citet{Kellermann_2016}, reveal that quasars (or QSOs) do not exhibit a strictly bimodal distribution in radio-loudness. Instead, there appears to be a continuous distribution with a broad overlap between the RL and RQ populations. In Radio Loud (RL)--AGNs, the radio emission predominantly arises from the jets, whereas in Radio Quiet (RQ)--AGNs, the origin of radio emission is not well understood and likely differs from one object to another. \citet{Panessa2019} discusses the proposed models for physical processes driving radio emission in RQ-AGNs, including emission due to AGN activity like small-scale jets, free-free emission from the AGN core, and accelerated charged particles at relativistic speeds in the corona. However, sometimes the radio emission from RQ-AGNs can be dominated by strong star formation in the host galaxy \citep{Padovani_2011,Condon_2013}.

The exact nature of how radio outflows are connected to accretion disk properties in AGNs remains unknown. 
Thus, understanding the synergy between accretion and outflows in the neighborhood of black-holes is important to our knowledge of how the latter impact their environments. 
For example, based on the strong observational correlations between optical and radio powers, a connection between the disk accretion rate and the production of radio jets has been suggested by \citet{1999MNRAS.309.1017W} for radio galaxies and by \citet{2001ApJ...555..650H} for Seyfert-I nuclei. \citet{2000A&A...356..445B} correlated the Roentgensatellit (ROSAT) All-Sky Survey (RASS, \citet{rosat1999A&A...349..389V}) and the Very Large Array - Faint Images of the Radio Sky at Twenty-Centimeters (VLA-FIRST, \citet{becker1994vla}) catalog and have shown that there is some correlation between the X-ray 2 keV luminosity and the radio 5 GHz luminosity for bright AGNs and quasars. \citet{Merloni2003} and \citet{2006A&A...456..439K} studied the correlations between the radio luminosity ($L_R$), X-ray luminosity ($L_X$), and black-hole mass ($M_{BH}$) for a sample of 100 AGNs and demonstrated that the sources define a \enquote{Fundamental Plane of black-hole Activity} (FP correlation). \citet{Payaswini2015} and \citet{Payaswini2018} found a similar correlation between the 1.4 GHz radio frequency emission and the X-ray emission / [O III] line emission from AGNs. However, follow-up studies such as \citet{Fischer_2021} and \citet{2024A&A...689A.327W} report deviations from the FP correlation among AGNs, which appear to depend on factors such as the accretion rate, AGN classification, and the nature of their radio emission. \citet{10.1093/mnras/staa1411, 10.1093/mnras/stab1406} investigated the previously examined $L_{2 keV} - L_{2500\mathring{\mathrm{A}}} - L_{5 GHz}$ relation \citep{Miller_2011} for radio-loud and radio-intermediate quasars (RLQ and RIQ) for a large sample of RL quasars, revealing connections between the disk, jet, and corona. 

Weak-Line Quasars (WLQ) are identified as a subclass of AGNs that typically exhibit weak high ionization emission-lines or the absence of emission-lines \citep{1999ApJ...526L..57F, 2001AJ....121...31F,2001AJ....122..503A, Collinge_2005, Fan_2006,2010ApJ...721..562P, Diamond-Stanic_2009,2014A&A...568A.114M}. The equivalent widths of high-ionization emission-lines like Ly$\alpha$ + NV and CIV are observed to be $\leq$ 15.4 $\mathring{\mathrm{A}}$ and $\leq$ 10 $\mathring{\mathrm{A}}$ in their rest frame UV spectra, respectively. In general, WLQs are radio-quiet AGNs and exhibit distinctive X-ray characteristics \citep{ Wu_2012,Luo_2015}. 
Approximately half of the WLQs exhibit weaker X-ray emission than predicted by the well-established correlation between X-ray ($L_{2\ \mathrm{keV}}$) and UV ($L_{2500\ \mathring{\mathrm{A}}}$) luminosities observed in quasars \citep{1986ApJ...305...57T, Brandt_2000,2006AJ....131.2826S, 2007ApJ...665.1004J,2007ApJS..173....1L,2007ApJ...663..103L, 2016ApJ...819..154L}. In contrast, the remaining half of the WLQs, not X-ray faint, exhibit steep X-ray spectra, signifying accretion occurring at elevated Eddington ratios or similar to that of normal quasars \citep{Wu_2011, Wu_2012, Luo_2015, ni2018connecting, ni2022sensitive}. Thus, WLQs exhibit a broad spectrum of X-ray luminosities.  

Several models have been proposed to explain the peculiar emission-line/X-ray properties of WLQs and their possible connection. For instance, \citet{2010ApJ...722L.152S}, \citet{2010MNRAS.404.2028H} and \citet{10.1093/mnras/stac3689} have proposed a model featuring an anemic broad emission-line region (BELR) or an underdeveloped BELR with a significant deficit of line-emitting gas, in contrast to the previously suggested scenario by \citet{2007ApJS..173....1L, 2007ApJ...663..103L}, which involved a supermassive black-hole (SMBH) with a large black-hole mass accreting at an extremely high Eddington ratio. Several studies, including \citet{Diamond-Stanic_2009}, \citet{2010ApJ...721..562P}, \citet{Luo_2015}, \citet{ni2018connecting}, \citet{ni2022sensitive} and \citet{Ha_2023}, suggest that WLQs are radio-quiet AGNs that accrete at high Eddington rates. A high black-hole mass system with a cold accretion disk was proposed by \citet{2011MNRAS.417..681L}. \citet{Shemmer_2015} studied a small sample of WLQs in the context of the modified Baldwin Effect and found that WLQ line properties do not solely depend upon the accretion Eddington ratio. Multiwavelength studies of WLQs by \citet{Wu_2011, Wu_2012, Luo_2015, ni2018connecting, ni2022sensitive} suggest an intrinsically obscured continuum. The Thick inner-accretion Disk and Outflow (TDO) model \citep{Luo_2015, ni2018connecting} offers a suitable explanation of the observed X-ray and emission-line properties of WLQs by proposing the presence of a puffed-up inner accretion disk obscuring the high-energy photons from reaching the BLR, giving rise to the observed optical spectra with blue-shifted weak UV emission-lines.

In the year 2020, \cite{ni2020extreme} reported an event characterized by an exceptional increase in X-ray intensity of a non-Broad Absorption Line (non-BAL) WLQ, SDSS J153913.47+395423.4 (hereafter SDSSJ1539+3954) in its Chandra observations. The quasar, at redshift $z\approx 1.935$, underwent a $\gtrsim$20-fold increase in soft X-ray flux (0.5--2 keV) from 2013 to 2019 (termed as \enquote{X-ray normal epoch}), followed by a $\sim$9-fold drop over a short period of $\lesssim$8 months in 2020, as reported by \citet{ni2022sensitive}. While the relationship between X-ray and UV/optical emissions in AGNs has been extensively studied, no significant changes were observed in the CIV line emission from the quasar. The object has a bolometric luminosity $L_{bol}=1.5 \times 10^{47} erg s^{-1}$, which makes it one of the most luminous WLQs to show such remarkable variability in X-ray.

The radio emission properties of WLQs remain relatively unexplored, particularly in comparison to their well-documented UV, optical, and X-ray characteristics. While numerous studies have established strong correlations between X-ray and radio emissions in typical quasars, it is unclear whether these relationships hold for WLQs, especially given their unusual emission-line strengths and diverse X-ray properties. In this context, investigating the radio emission of SDSSJ1539+3954 is particularly interesting, as the source has exhibited remarkable X-ray variability.  In this study, we analyze SDSSJ1539+3954 across multiple radio frequency bands, aiming to explore any correlation between the observed variability in X-ray and radio emissions of this quasar, particularly investigating the presence of radio variability and the mechanisms underlying radio emission in this radio-quiet source. 

A description of the X-ray properties of the source and radio/optical data analysis is given in the section \ref{sec:data}. Section \ref{sec:results} includes results obtained from the radio/optical data analysis and their interpretations. We discuss the implications of observed properties of the source in the section \ref{sec:discussion}. The main conclusions are summarized in the section \ref{sec:conclusion}. For this work, we assume a flat cosmological model of the universe and adopt the following parameters: H$_0$ = 69.6 km/s/Mpc, $\Omega_M$ = 0.286, and $\Omega_{vac}$ = 0.714.

\section{Observations and data analysis}\label{sec:data}
\subsection{Archival X-ray data}
\citet{ni2022sensitive} observed the quasar SDSSJ1539+3954 in the X-ray energy band of 0.5-8.0 keV using the Advanced CCD Imaging Spectrometer spectroscopic array \citep{garmire2003advanced} on board the Chandra satellite in very faint (VFAINT) mode in 2013, 2019 and 2020. They analyzed the data sets in energy bands 0.5-8.0 keV (full-band), 0.5 - 2.0 keV (soft-band), and 2.0-8.0 keV (hard band) separately. The source was undetected in 2013 and exhibited an exceptional increase in X-ray intensity in the soft-band X-ray region in the 2019 observation. Between 2013 and 2019, a $\gtrsim 20$ fold increase in X-ray flux was observed from the object. The soft X-ray flux from SDSSJ1539+3954 declined again in 2020 by a factor of $\ge 9$ in a short period of $\lesssim$8 months, falling below the detection threshold again. 
Tables 1 and 3 in \citet{ni2020extreme} and \citet{ni2022sensitive}, respectively, contain a comprehensive list of reported source parameters for each Chandra observation of the source, including source counts, observed spectral index, and observed fluxes from the source in both soft (0.5 - 2.0 keV) and hard (2.0 - 8.0 keV) X-ray bands. The source was detected only in 2019 during the X-ray brightening phase, remaining undetected in all previous and subsequent epochs. The Chandra full-band (0.5--8.0 keV) spectrum obtained during the X-ray bright phase in 2019 was observed to follow an effective power-law photon index of $\Gamma_{eff} = 2.0 \pm 0.4$.

\subsection{Radio Observations}
To assess the radio spectral properties of the quasar SDSSJ1539+3954 across different frequencies and investigate any potential variability, specifically in the range of 340 MHz to 10 GHz, we utilize both archival data and targeted observations of the source. The above-mentioned frequency bands are selected to examine the total emission, including both core and extended emission, from the source. We present the information obtained from radio surveys in the section \ref{subsec:radioarchive}. Details of radio observations of the source and data analysis are discussed in sections \ref{subsec:GMRT_analysis} and \ref{subsec:vla_analysis}.

\subsubsection{Archival Radio data}\label{subsec:radioarchive}
In order to analyze the behavior of the source across different radio frequency bands, we assessed the availability of archival data. The VLA-FIRST survey at 1.4 GHz \citep{1995ApJ...450..559B} and the Very Large Array Sky Survey (VLASS) \citep{2020PASP..132c5001L} at 3 GHz provide quick-look images for the radio sky at frequencies 1.4 GHz and 3 GHz conducted from the years 1993 to 2011 and 2017 to 2023, respectively. Radio intensity images for the source SDSSJ1539+3954 obtained from the VLA-FIRST and VLASS surveys undertaken in 1994 and 2017 to 2023 do not show radio emission detection for this source with a background rms of about 140 $\mu$Jy/beam and 120 $\mu$Jy/beam for the VLA-FIRST and VLASS surveys, respectively. The non-detection of the source in the available archival data at around 1.4 GHz and 3 GHz can be attributed to faint radio emission from the source and the sensitivity limits of both surveys.

\subsubsection{GMRT observations }\label{subsec:GMRT_analysis}
We obtained radio interferometric data from the upgraded Giant Meterwave Radio Telescope (GMRT) in the years 2020, 2022, and 2024, following the reported X-ray variability epochs. We observed the source in the years 2020 (Project ID: ddtC119 and ddtC140, PI: Vivek M), 2022 (Project ID: 41$\_$062, PI: Vivek M) and 2024 (Project ID: 46$\_$028, PI: Ayushi Chhipa) using three frequency bands of the GMRT, namely Band-3 (250-500 MHz), Band-4 (550-850 MHz) and Band-5 (1050-1450 MHz) using the full 29 antennae configuration. GMRT observations conducted in 2020 are quasi-simultaneous with the X-ray observation obtained from Chandra. The datasets were recorded in total intensity mode with 2048 frequency channels, spanning a bandwidth of 400 MHz in bands 5 and 4, and 200 MHz in band 3, using an integration time of 10 s. The total observation time was about 3 hours for all the frequency bands. 3C286 was used as the flux and bandpass calibrator for GMRT observations conducted in all frequency bands. J1602+334(or J1602+3326) and J1609+266 were used as phase calibrators for Bands-4,5 and Band-3 GMRT observations, respectively.

We use the Common Astronomy Software Application (CASA) software, developed by \citet{bean2022casa}, to reduce and analyze the radio data collected from the radio observations. The initial flagging of the data and removal of radio frequency interference (RFI) was done using standard CASA tasks. Specifically, we utilized the TFCROP mode of the FLAGDATA task for initial flagging, followed by manual flagging of corrupted channel bins and time bins using the MANUAL mode in the FLAGDATA task. The \citet{perley2017accurate} scale was adopted to set the flux density of the flux calibrator 3C286.
The basic calibration rounds, including phase, delay, bandpass, and complex gain calibrations, were carried out according to standard procedures.
We extracted the calibrated target source visibilities using the SPLIT task in CASA. This extraction was performed with averaged spectral channels to reduce data volume without introducing bandwidth smearing.

The Stokes-I images of the target were generated using the TCLEAN task in CASA, employing the multi-term multi-frequency synthesis (mtmfs) deconvolver as described by \citet{rau2011multi} and utilizing the Briggs weighting scheme. We use phase-only self-calibration along with multiple TCLEAN rounds to improve the image fidelity. 
The approximate beam sizes obtained for Band-3, Band-4, and Band-5 images are 6.5" $\times$ 5.3", 3.7" $\times$ 3.4" and 2.9" $\times$ 1.9" respectively. 
We utilize the Gaussian fitting method implemented in the IMFIT task in CASA to calculate the integrated flux density emission from the source. The flux density errors are estimated using the formula $\sqrt{(\sigma_{rms}+(\sigma_p S)^2}$ \citep{weiler1996radio,10.1093/mnrasl/slz061}, where $\sigma_{rms}$ is the rms noise of the radio images for the point source region obtained from the Gaussian fit, S is the source flux density, and $\sigma_p$ is the absolute percentage error in the flux density, which is assumed to be 10$\%$ (conservative for GMRT and VLA).

\begin{figure*}
\centering
    \hspace{-1.4cm}
    \begin{tabular}{ccc}
        \includegraphics[scale=0.33]{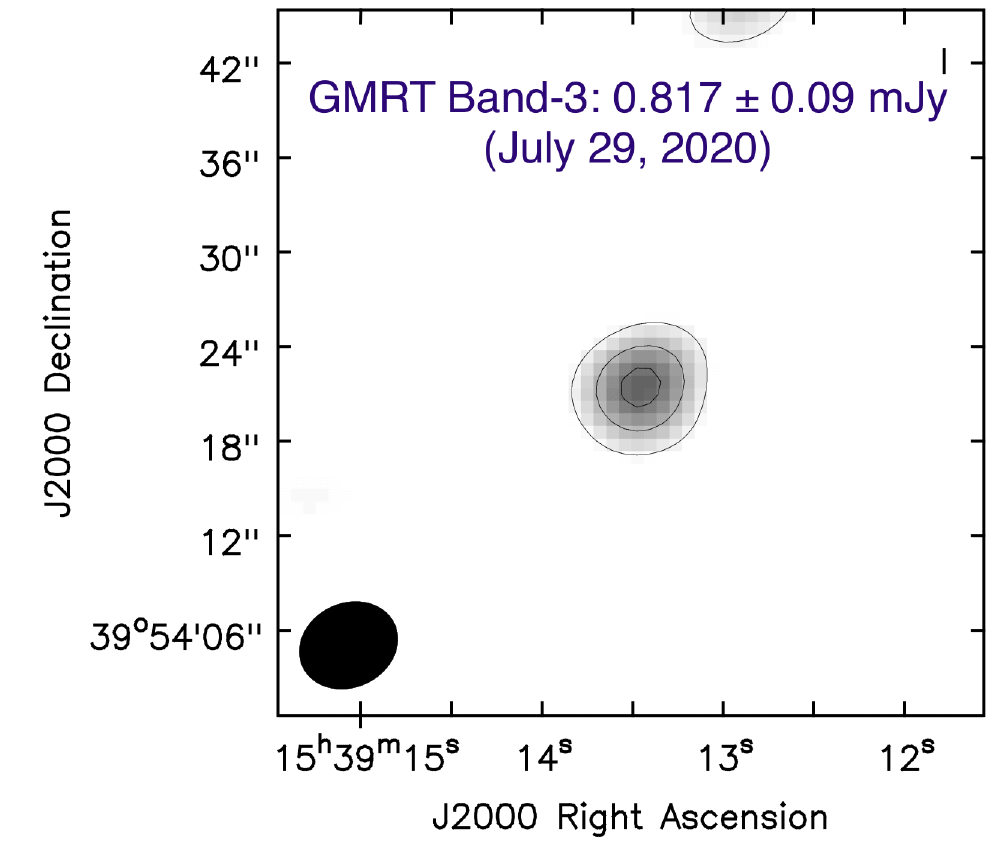} &   
       \includegraphics[scale=0.33]{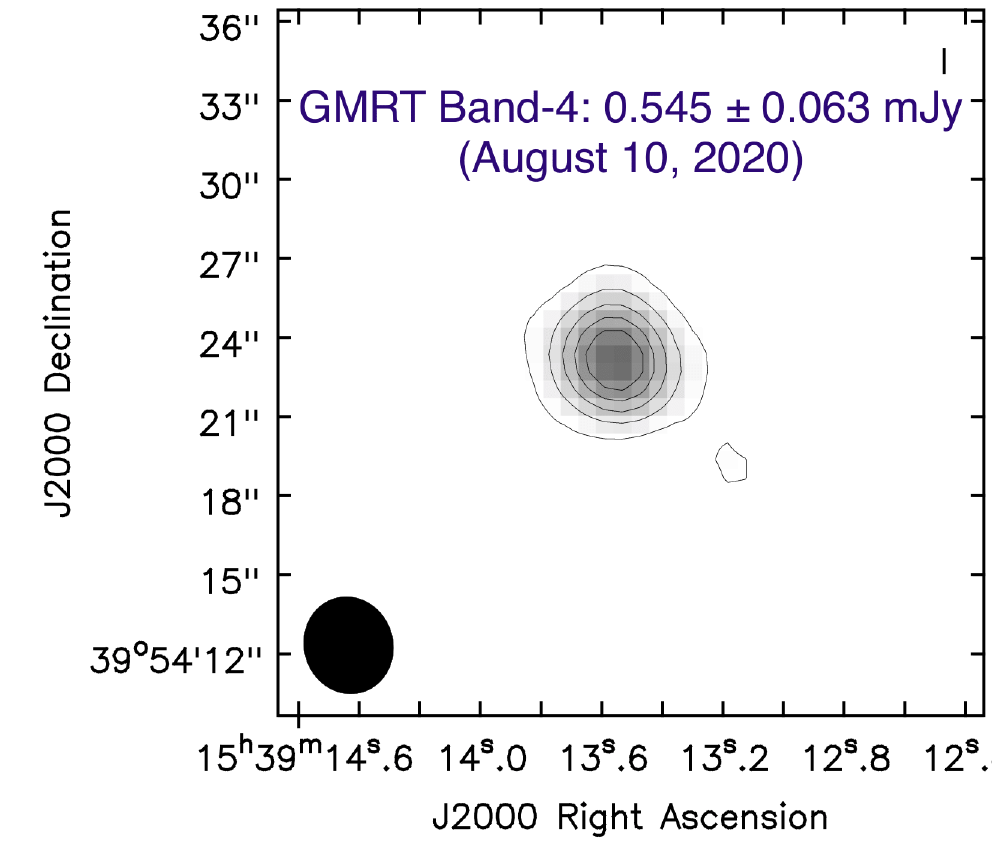}  &  
       \includegraphics[scale=0.33]{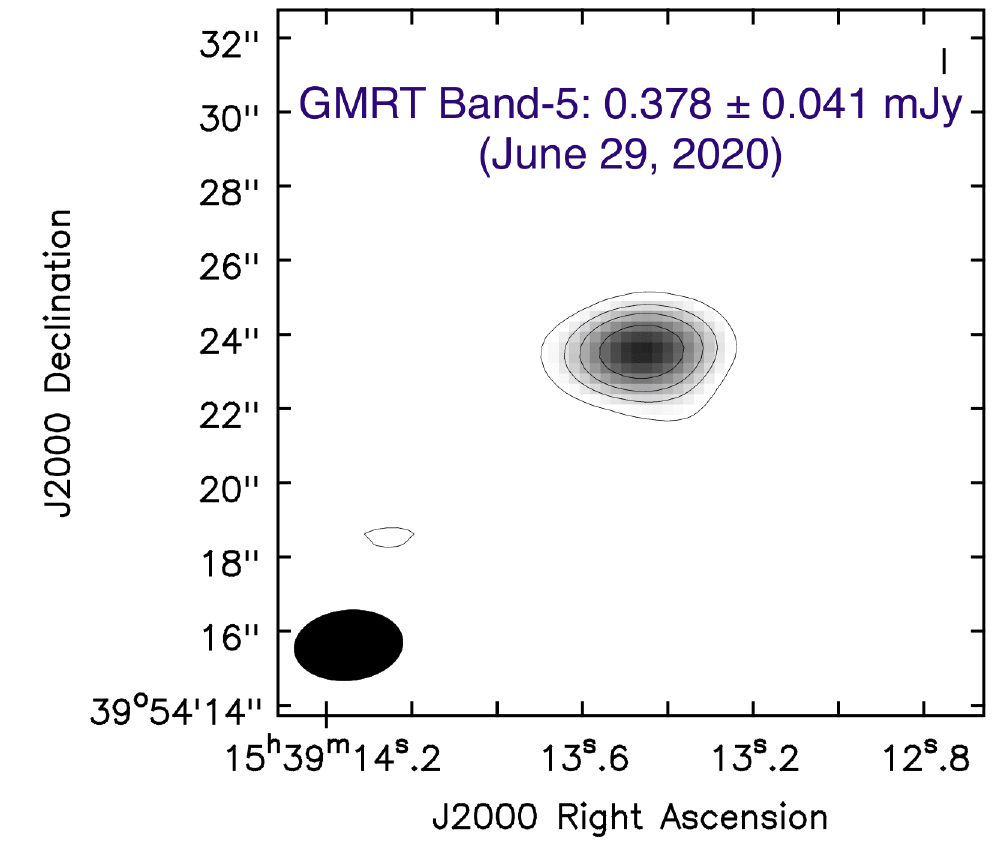}  \\
       \includegraphics[scale=0.33]{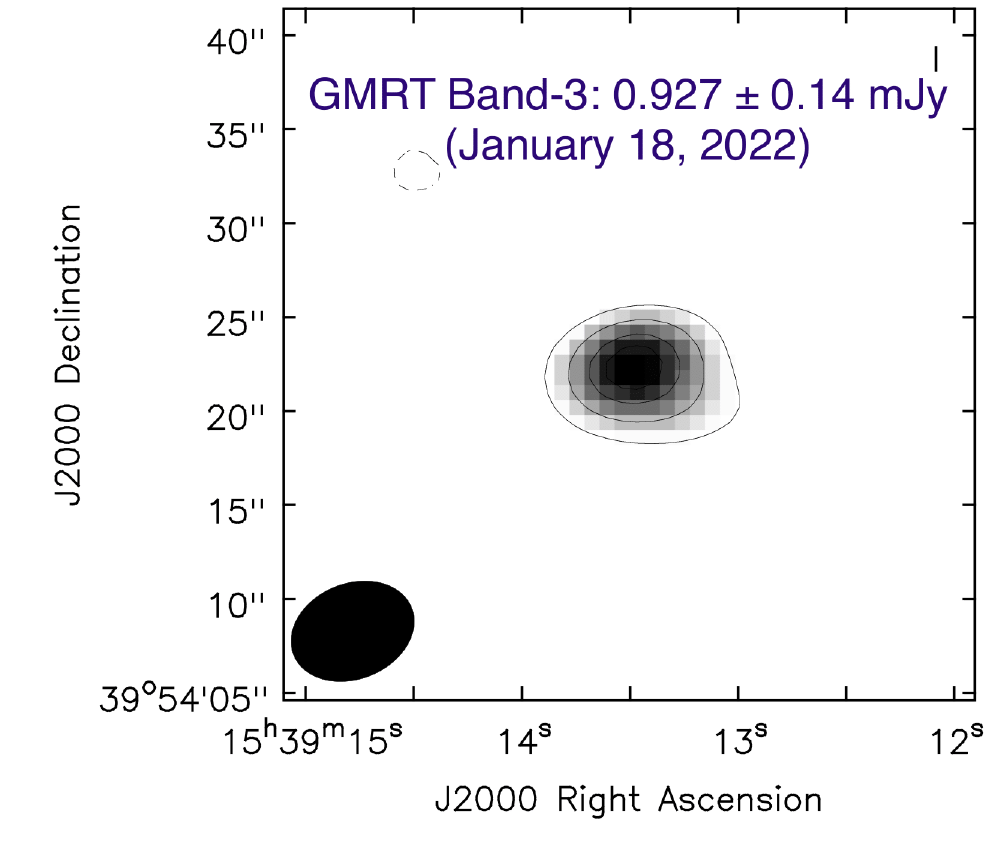} &   
       \includegraphics[scale=0.33]{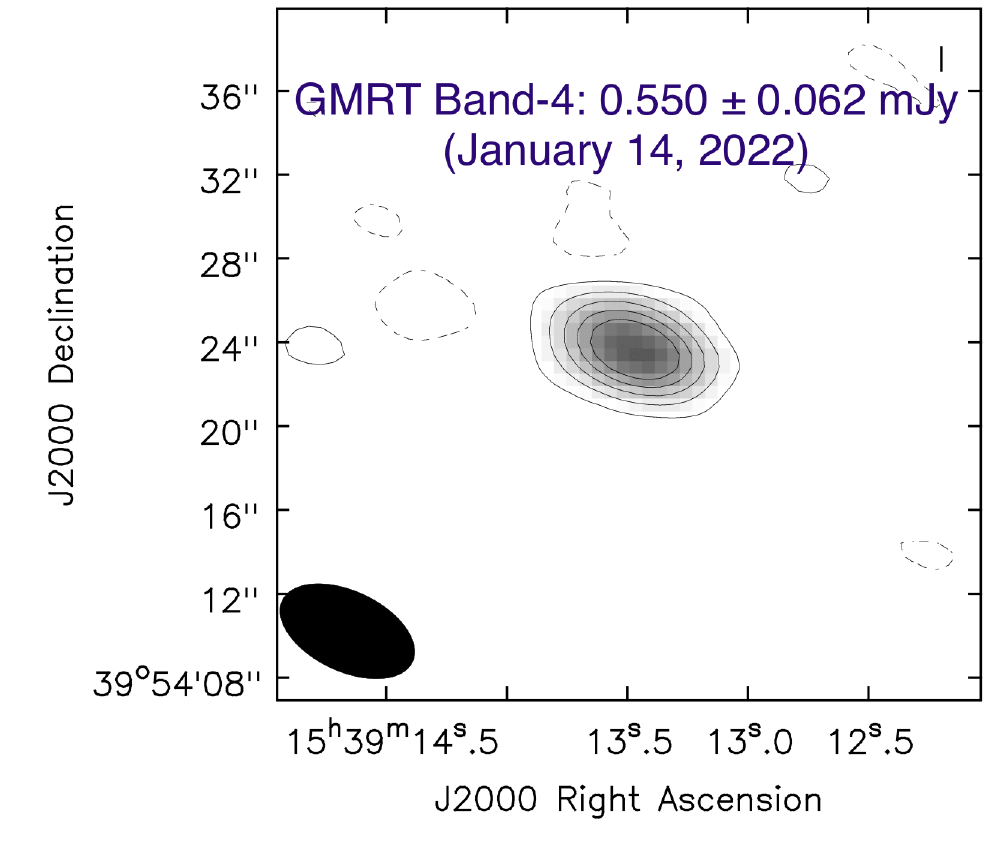}  &  
       \includegraphics[scale=0.33]{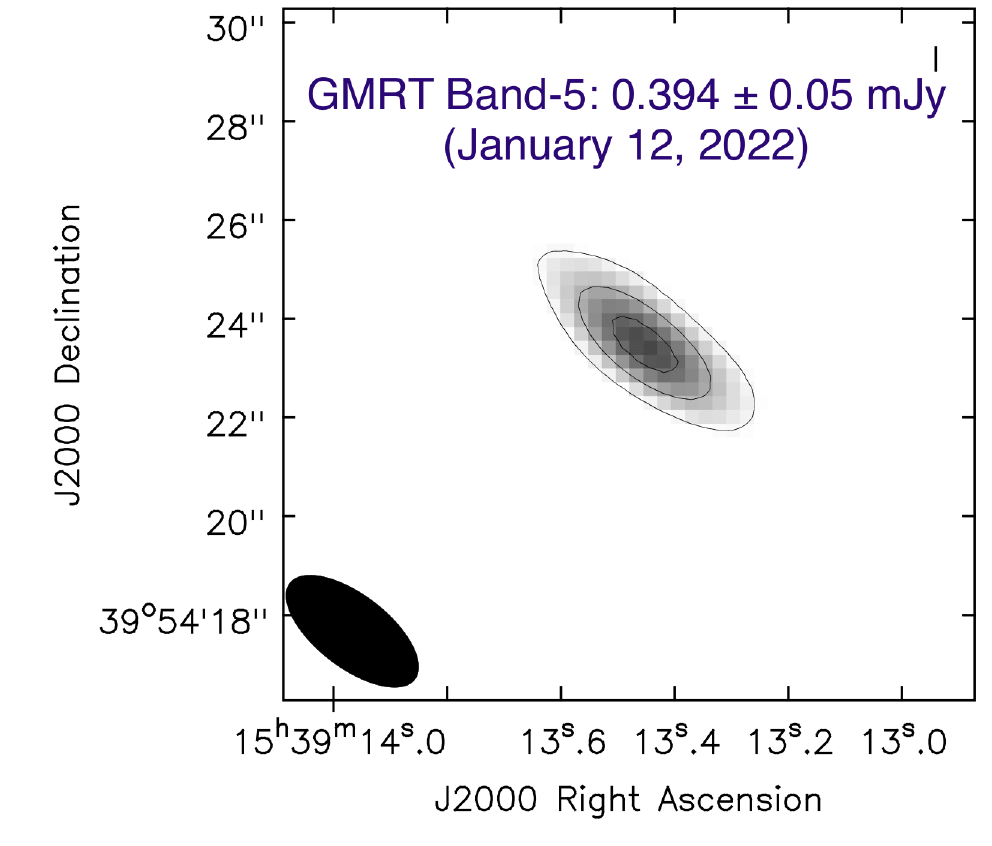}  \\
       &
       \includegraphics[scale=0.33]{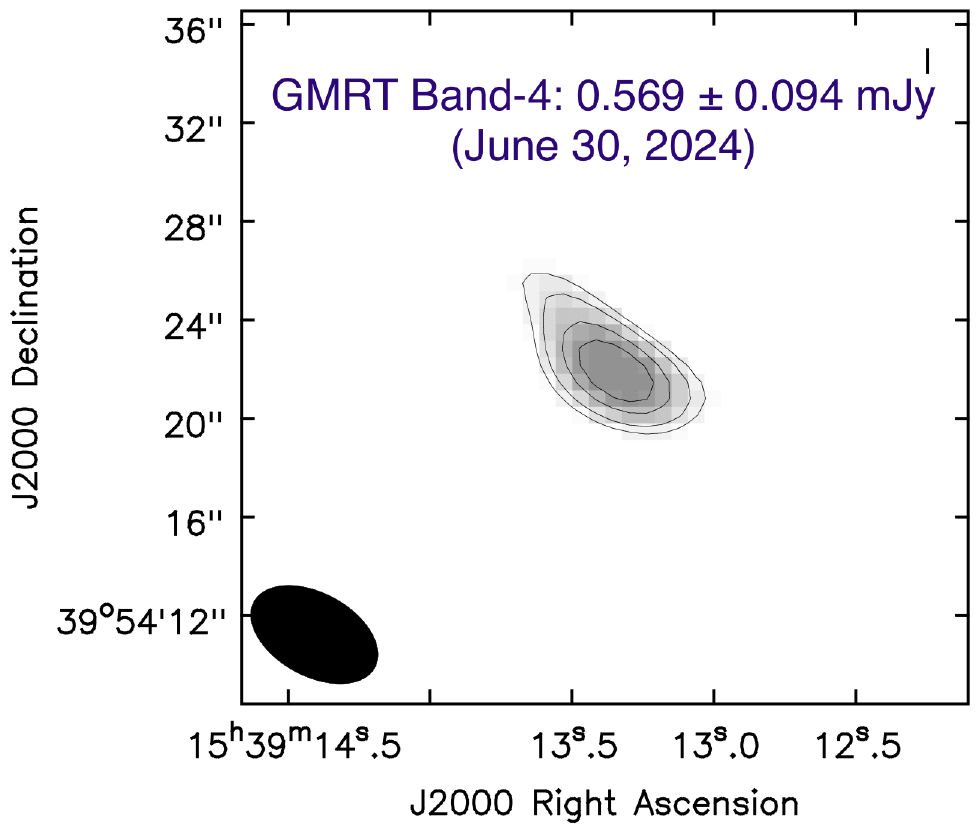}  &  
       \includegraphics[scale=0.33]{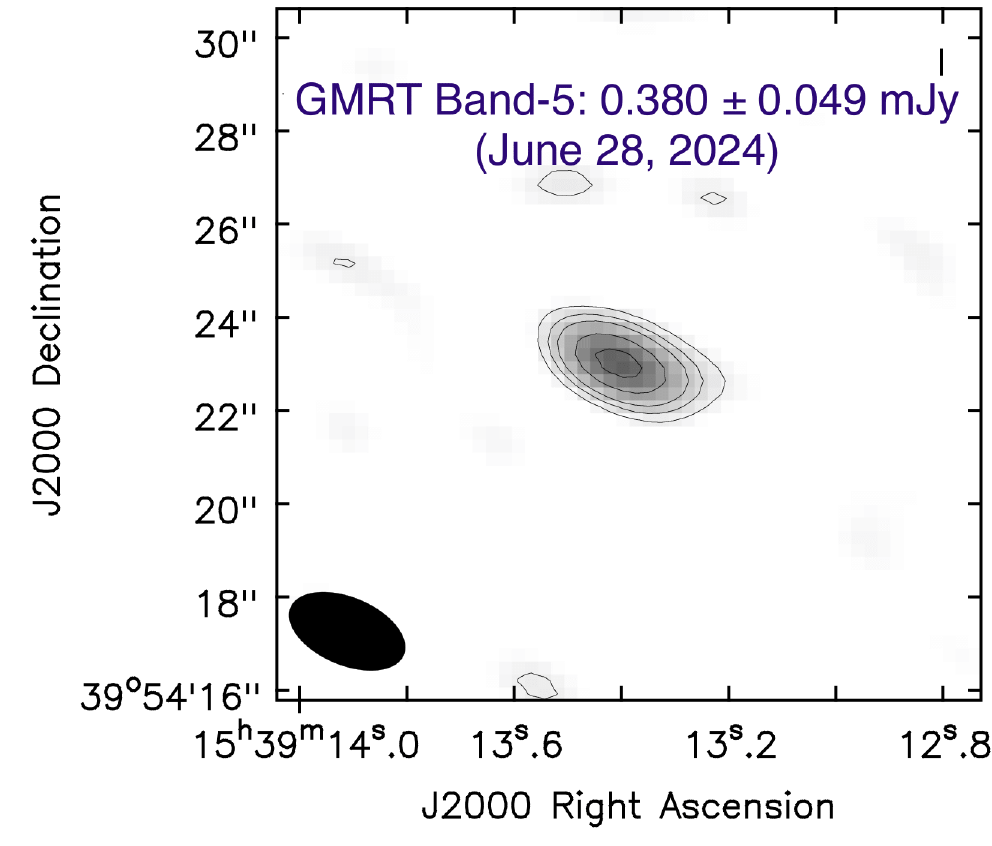}  \\
          
    \end{tabular}
    \caption{Flux density images with overplotted contours (dashed lines are used for negative contours) for the GMRT observations conducted in years 2020 (top row), 2022 (middle row), and 2024 (bottom row). The contour levels scale as: (image rms noise) $\times$ [-3, 3, 7, 9, 13, 15]. All the images are arranged in increasing order of frequency from left to right columns at 340 MHz, 750 MHz, and 1250 MHz, respectively. The solid ellipse at the bottom of every image represents the beam shape. The approximate beam sizes obtained for Band-3, Band-4, and Band-5 images are 6.5" $\times$ 5.3", 3.7" $\times$ 3.4" and 2.9" $\times$ 1.9" respectively. Respective flux density values for the source SDSSJ1539+3954 are mentioned within the image panels along with the corresponding frequency band and observation date.}
    \label{fig:GMRTimages}
\end{figure*}

Figure \ref{fig:GMRTimages} shows the radio intensity images for the source field in all the GMRT bands, Band-3 (250-500 MHz, left column), Band-4 (550-850 MHz, middle column),
and Band-5 (1050-1450 MHz, right column). The three-row panels show the radio images obtained in the year 2020 (epoch 1; top row), 2022 (epoch 2; middle row), and 2024 (epoch 3; bottom row), respectively. The beam shape, flux densities, and observation dates are noted in every image. The acquired flux densities for GMRT Bands-3,4,5 in 2020 are 0.817$\pm$0.090 mJy, 0.545$\pm$0.063 mJy and 0.378$\pm$0.041 mJy, respectively. Similarly, for the GMRT observations 2022 and 2024, the obtained flux densities for Bands-3,4,5 are: 0.927$\pm$0.140 mJy (2022); 0.550$\pm$0.062 mJy (2022) and 0.569$\pm$0.094 mJy (2024); and 0.394$\pm$0.050 mJy (2022) and 0.380$\pm$0.049 mJy (2024), respectively. The observed flux densities are also listed in Table \ref{tab:table-2020-GMRT} for epochs 1, 2, and 3, respectively.

\subsubsection{VLA observations} \label{subsec:vla_analysis}
We also obtained observations in the S (2.0--4.0 GHz), C (4.0--8.0 GHz), and X (8.0--12.0 GHz) bands with the Karl G. Jansky Very Large Array (VLA; Project ID: 22A-166, PI: Vivek M) in the A-configuration using all 27 antennas in 2022, to examine the broadband radio spectral behavior at higher frequencies and to probe the core emission of the source. For VLA observations, the data were recorded using a 3-bit sampler in bands C and X, covering a 4 GHz bandwidth, and an 8-bit sampler in band S, covering a bandwidth of 2 GHz. The source J1602+334(or J1602+3326) was used as the phase calibrator for VLA observations.

The initial flagging and calibration for the data obtained from the VLA observations were conducted using the VLA Calibration PIPELINE in CASA (version 6.2.1.7).
We extracted the calibrated visibilities of the target source using the SPLIT task in CASA, and the Stokes-I images of the target were generated using the TCLEAN task in CASA, employing the mtmfs deconvolver utilizing the Briggs weighting scheme. The approximate beam sizes obtained for Band-S, Band-C, and Band-X images are 1.1" $\times$ 0.54", 0.51" $\times$ 0.3", and 0.33" $\times$ 0.23" respectively. 

\begin{figure*}[h!]
    \centering
    \hspace{-0.5cm}
      \includegraphics[scale=0.33]{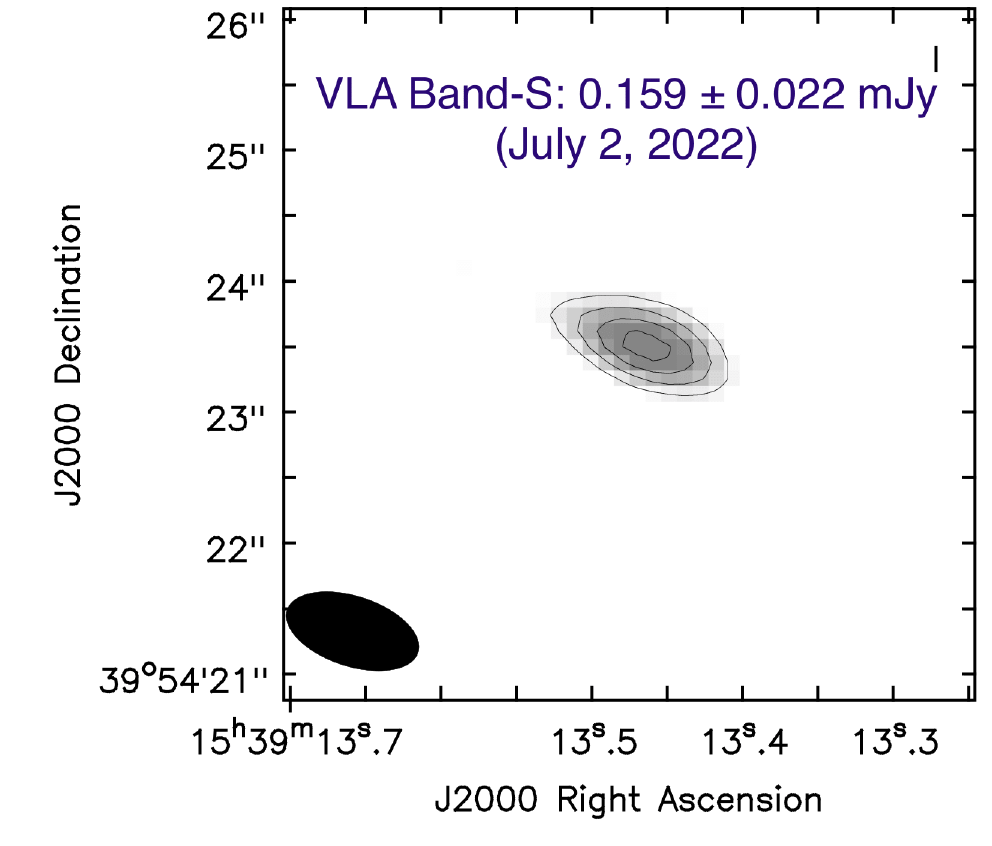}
        \includegraphics[scale=0.33]{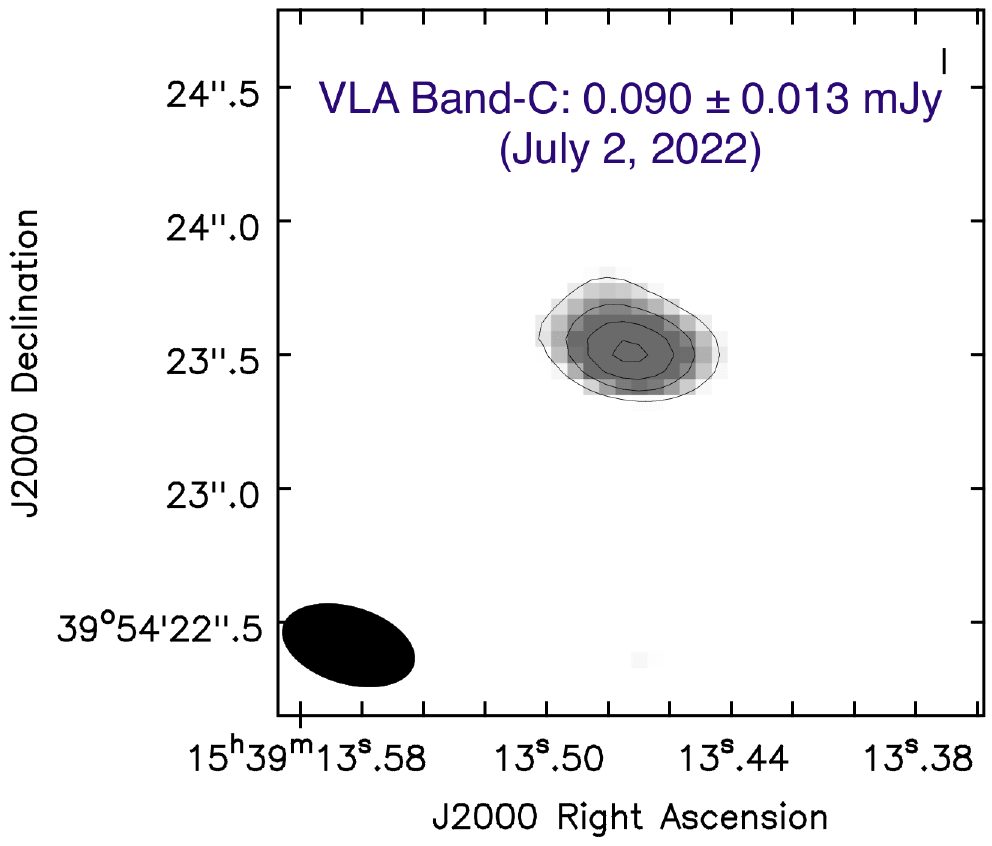}
        \includegraphics[scale=0.33]{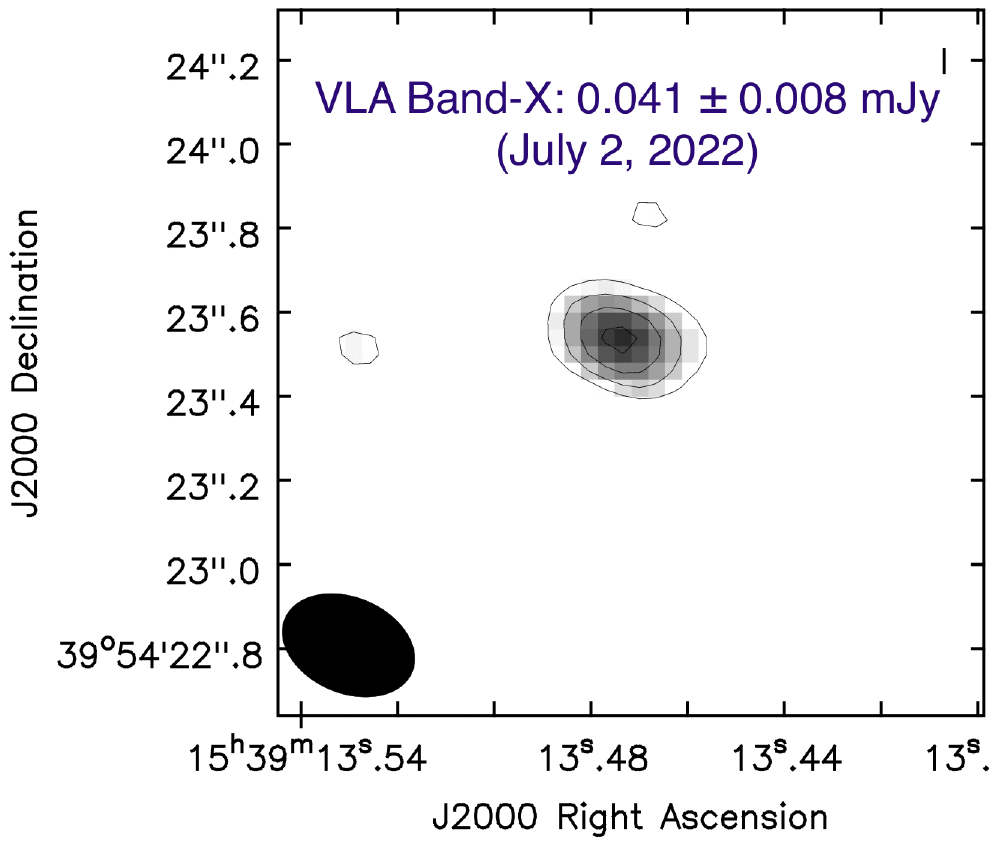}
    \caption{Flux density images with overplotted contours for the VLA observations conducted in 2022. flux density image of the source at frequency 2999 MHz (left panel), 5999 MHz (middle panel), and 9999 MHz (right panel). The contour levels scale as: (image rms noise) $\times$ [-3, 3, 7, 13, 15]. The solid ellipse at the bottom of every image plane represents the beam shape. The approximate beam sizes obtained for Band-S, Band-C, and Band-X images are 1.1" $\times$ 0.54", 0.51" $\times$ 0.3", and 0.33" $\times$ 0.23" respectively. Respective flux density values for the source SDSSJ1539+3954 are mentioned within the image planes along with the corresponding frequency band and observation date.}
    \label{fig:vlaimages}
\end{figure*}

In the analysis of the VLA data as well, we introduced a systematic error of $10\%$ in bands S, C, and X to accommodate calibration uncertainties, alongside the Gaussian fit error similar to the GMRT data analysis. Figure \ref{fig:vlaimages} shows the radio intensity images for the source field in the three VLA bands, Band-S (2.0-4.0 GHz; left panel), Band-C (4.0-8.0 GHz; middle panel), and Band-X (8.0-12.0 GHz; right panel) as observed in the year 2022 (epoch 2). The beam shape, flux densities, and observation dates are noted in every image. Flux density of 0.159 $\pm$ 0.022 mJy, 0.090$\pm$0.013 mJy, and 0.041$\pm$0.008 mJy are obtained for the VLA Bands-S, C, and X. The observed flux densities are listed in Table \ref{tab:table-2020-GMRT} along with the GMRT observed flux densities in the year 2022.

\begin{table*}[h!]
    \centering
    \hspace{-2cm}
    \begin{tabular}{ccccccc}
    \hline
        Observation &Observation  & Observation  &Duration of & Image Frequency& Observed  & rms noise \\
         Epoch&Date & Band& observation& & flux density &  \\
        &&& (in Min.)&(in MHz) & (in mJy) & (in mJy/beam)\\
         \hline
         &29-07-2020& GMRT Band-3 & 133.0&340 & 0.817 $\pm$ 0.090 & 0.056 \\
        1& 10-08-2020& GMRT Band-4 & 120.0& 750 & 0.545 $\pm$ 0.063 & 0.021 \\
         &29-06-2020& GMRT Band-5 & 105.0& 1260 & 0.378 $\pm$ 0.041 & 0.014\\
         \hline
        & 18-01-2022&GMRT Band-3 & 107.0& 340 & 0.927 $\pm$ 0.140 & 0.048 \\
         &14-01-2022 &GMRT Band-4& 130.0& 750 & 0.550 $\pm$ 0.062 & 0.023 \\
      2  & 12-01-2022 &GMRT Band-5& 104.5& 1260 & 0.394 $\pm$ 0.050 & 0.027 \\
        & 02-07-2022 &VLA Band-S& 16.8& 2999 & 0.159 $\pm$ 0.022 &0.015\\
       &  02-07-2022 &VLA Band-C& 16.8& 5999 & 0.090 $\pm$ 0.013&0.009\\
        & 02-07-2022 &VLA Band-X& 16.8 & 9999 & 0.041 $\pm$ 0.008 &0.006\\
         \hline
     3    &30-06-2024& GMRT Band-4 & 64.5& 650 & 0.569 $\pm$ 0.094 & 0.025 \\
         &28-06-2024& GMRT Band-5 & 52.0& 1260 & 0.380 $\pm$ 0.049 & 0.02\\
         \hline
    \end{tabular}
    \caption{Observed flux density for the source SDSSJ1539+3954 in GMRT Band-3, 4 and 5 and VLA Band-S, C and X observations in 2020 (epoch 1), 2022 (epoch 2) and 2024 (epoch 3). Errors in flux densities are derived by adding a 10$\%$ systematic uncertainty with background rms in quadrature.}
    \label{tab:table-2020-GMRT}
\end{table*}

\subsubsection{Optical spectra and light curves}
To evaluate how the strong X-ray fluctuations impact the optical continuum and line emission properties of the source, we obtained its spectra from the SDSS-DR16 catalog \citep{2020ApJS..250....8L}, which includes observed spectra of the source from 2002, 2012, and 2017. Additionally, we used the 2m Himalayan Chandra Telescope - Himalayan Faint Object Spectrograph Camera (HCT - HFOSC) to observe the source following the X-ray brightening and dimming episodes in 2020 and 2021, capturing its spectrum. HFOSC is an imaging cum spectrograph instrument used for spectroscopy. 
We have obtained spectra for the source SDSSJ1539+3954 using the Grism-7 covering the wavelength regions from 3000 - 8000 $\mathring{\mathrm{A}}$. A FeAr lamp was used for the wavelength calibration for the observation, and a Halogen lamp as a flat field. 

We analyzed the data using the standard packages of the IRAF software \citep{10.1117/12.968154}. After performing the initial bias, flat, and cosmic ray correction, we extracted the source spectrum using the DOSLIT package. Furthermore, to check for optical continuum variability in the source, we acquire the light curve from the Zwicky Transient Facility (ZTF-IPAC) in the g-band as well. The observed properties of the source in the optical spectra and ZTF g-band light curve are discussed in Section \ref{sec:results}.

\section{Results}\label{sec:results}
\begin{figure*}[ht!]
    \centering
    \hspace{-0.38cm}
    \includegraphics[scale=0.34]{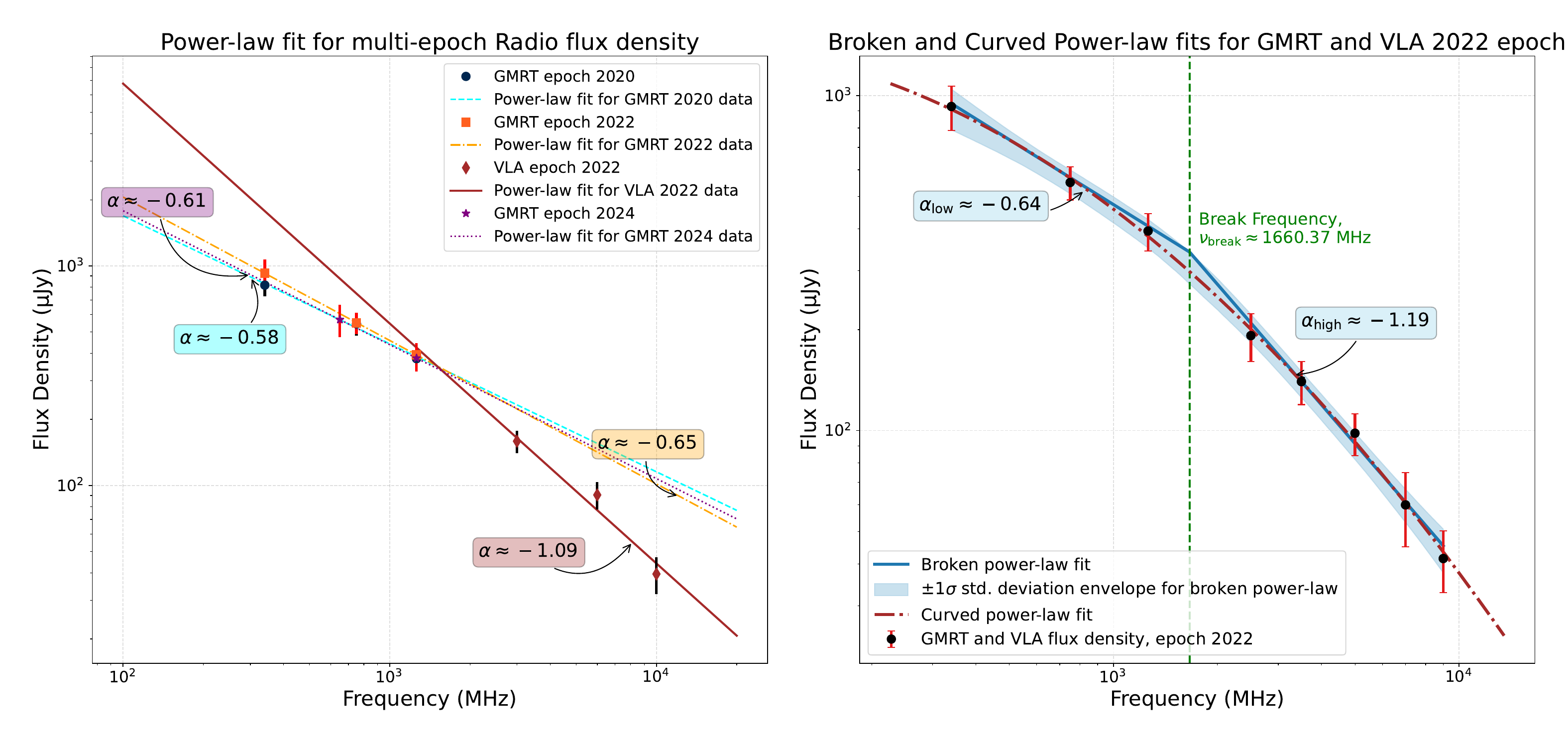}
    \caption{Left panel: power-law fits for the radio flux densities obtained from GMRT (300 MHz--1.26 GHz) and VLA (3--10 GHz) observations. The cyan dashed line represents the GMRT epoch 2020 fit with spectral index $-0.58$, the yellow dot-dashed line for the GMRT epoch 2022 with spectral index $-0.65$, the solid brown line for the VLA epoch 2022 with spectral index $-1.09$ and the dotted magenta line for the GMRT epoch 2024 with spectral index $-0.61$. GMRT epoch 2020 flux densities are shown as circles (blue), epoch 2022 flux densities are shown as squares (orange), and epoch 2024 flux densities as stars (magenta). The VLA epoch 2022 flux densities are shown as diamonds (brown). \\ Right panel: broken power-law fit (blue solid line) and curved power-law fit (brown dot-dashed line) for the flux densities obtained from the GMRT observation epoch 2022 and sub-bands from the VLA observation epoch 2022. The spectral index changes from $\alpha_{low} \approx -0.64$ to $\alpha_{high} \approx -1.19$ at the break frequency $\nu_{break} \sim 1660.37$ MHz. The shaded region shows the 1-$\sigma$ deviation in the fit parameters over several bootstrap iterations to fit the broken power-law. Similarly, the curved power-law fit provides a spectral index of $\alpha = -0.78 \pm 0.08$ which steepens at a rate of $2\beta$ per decade in frequency near 1 GHz, where $\beta = -0.31 \pm 0.13$.}
    \label{fig:radio-plots}
\end{figure*}

Figures \ref{fig:GMRTimages} and \ref{fig:vlaimages} present the radio intensity images obtained for the source from the GMRT and VLA observations conducted in 2020, 2022, and 2024, spanning the frequency range from 340 MHz to 10 GHz. Unlike the VLA-FIRST and VLASS epoch observations, the source is well detected across the full frequency range of 340 MHz to 10 GHz. The structure of the source remains unresolved across all frequencies, appearing point-like in all images. This unresolved nature can be attributed to the combination of its high redshift ($z\sim 1.93$) and the baseline limitations of GMRT and VLA.  We achieve a maximum resolution of approximately 0.23" with the VLA X-band (10 GHz) observation of the source, corresponding to a spatial scale of up to 2 kpc in the source rest frame. The flux densities measured across all three GMRT observation epochs are consistent within their respective uncertainties, indicating no significant radio variability over six years.

To characterize the spectral properties of the radio emission from the source, we use a power-law model of the form,
\begin{equation}
    F_\nu = A \nu^{\alpha}
\end{equation} applied to the flux densities obtained from GMRT and VLA observations at different frequencies. Where $F_\nu$ is the flux density obtained at the frequency $\nu$, and $A$ and $\alpha$ are the estimated scale factor and spectral index, respectively. We use the non-linear least squares optimization method to estimate the radio emission spectral parameters.
The measured flux densities for all the epochs of the GMRT observations are consistent with emissions that are non-thermal in nature with spectral indices, $\alpha = -0.58 \pm 0.11$, $\alpha = -0.65 \pm 0.15$ and $\alpha = -0.61 \pm 0.3$ in the frequency range of 300 MHz to 1.26 GHz. On the other hand, the flux densities in the VLA frequency range of 3 GHz to 10 GHz exhibit an even steeper spectral index of $\alpha = -1.09 \pm 0.16$. Thus, the spectral index changes and gets steeper for frequencies above 3 GHz, suggesting the aging of synchrotron-emitting particles. The left panel of Figure \ref{fig:radio-plots} shows the obtained flux densities for all the frequencies from both GMRT and VLA and the respective power-law fits.

To characterize the energy loss time scales for an optically thin synchrotron radiation emitting plasma \citep{2022MNRAS.515..473P} and the change in spectral index over the full range of frequencies, we use broken power-law and curved power-law of the forms:
\begin{equation}\label{eq:broken}
F_\nu = A \begin{cases}
\left( \dfrac{\nu}{\nu_{\text{break}}} \right)^{\alpha_{\text{low}}}, & \text{for } \nu < \nu_{\text{break}} \\
\left( \dfrac{\nu}{\nu_{\text{break}}} \right)^{\alpha_{\text{high}}}, & \text{for } \nu > \nu_{\text{break}}
\end{cases}
\end{equation}
and \begin{equation}\label{eq:curved}
    F_\nu = A \  \nu^{(\alpha \ +\ \beta log\nu)}
\end{equation}
where the $\alpha_{low}$ and $\alpha_{high}$ represent the spectral index for frequencies lower and higher than the spectral break frequency, $\nu_{break}$, and $\beta$ represents the change in the spectral index ($\alpha$) over the full frequency range.
To obtain better constraints on the spectral break frequency, we subdivide the VLA observation bands--S, C, and X--into narrower sub-bands such that the resulting sub-band images achieve a signal-to-noise ratio of $\ge$5. Accordingly, the S-band (total bandwidth: 2 GHz) is divided into two sub-bands of $\sim$1 GHz each, and the C and X-bands (total bandwidth: 4 GHz each) into two sub-bands of 2 GHz each.

We obtain eight data points to fit the broken and curved power-law models. However, the number of data points is still too small to fit a five-parameter model and obtain a reliable break frequency estimate. Hence, to obtain the break frequency, we apply a parametric bootstrap approach, iterating 10,000 times by perturbing the measured sub-band flux densities with Gaussian noise based on their 1$\sigma$ uncertainties, and refitting the broken power-law model to each realization. This enables us to derive the median parameter values and their uncertainties from the resulting distribution of best-fit solutions. The right panel of Figure~\ref{fig:radio-plots} presents the resulting spectral indices, $\alpha_{\mathrm{low}} = -0.64 \pm 0.18$ and $\alpha_{\mathrm{high}} = -1.19 \pm 0.32$, with a spectral break at $\nu_{\mathrm{break}} \approx 1660.37$ MHz. The median best-fit broken power-law model is shown as a solid blue line, with the shaded region representing the 1$\sigma$ spread from the bootstrap fit realizations. We use the non-linear least squares optimization method to fit the curved power-law model. We obtain a spectral index of $\alpha = -0.78 \pm 0.08$, with a curvature parameter $\beta = -0.31 \pm 0.13$, corresponding to a spectral steepening rate of $2\beta = -0.62$ per decade around 1 GHz. This further translates to a steepening rate of approximately $0.87\beta = -0.27$ per GHz. The dot-dashed best-fit line in the right panel of Figure~\ref{fig:radio-plots} presents the best-fit obtained for the curved power-law model. The parameters derived from both the broken and curved power-law fits are consistent with each other within 3$\sigma$ error limits, indicating a steepening of the spectral index with increasing frequency. The reduced $\chi^2$ values of 0.37 and 0.16 were obtained for the broken and curved power-law fits, respectively, indicating that the curved power-law provides a better fit. We note, however, that these low reduced $\chi^2$ values suggest overfitting, which may result from the small number of data points and their relatively large uncertainties.

Assuming an injection spectral index of $\alpha_{\text{low}} = -0.64$, we estimate the radio luminosity to be $L_{\text{rad}} \approx 4.01 \times 10^{41}$ ergs/s \citep{1970ranp.book.....P}. The corresponding magnetic field strength is approximately 3.23 $\mu$G. Using this, the upper limit for synchrotron cooling timescale is estimated to be $\sim$0.56 Myr \citep{2016era..book.....C}, potentially indicating an absence of recent particle acceleration or renewed radio emission in the source.

\begin{figure*}[h!]
    \centering
    \hspace*{-1.13cm}
    \includegraphics[scale=0.43]{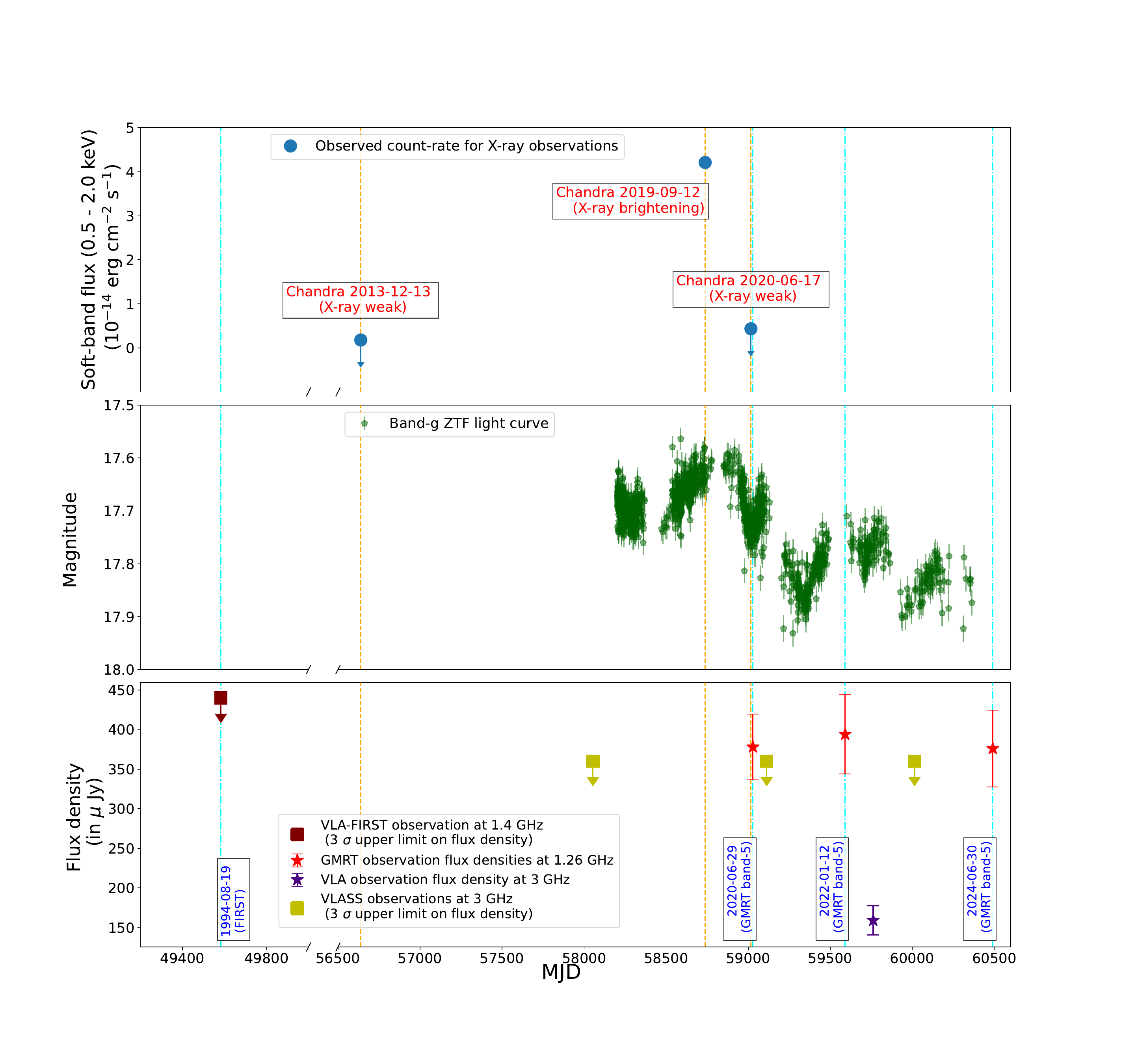}
    \caption{Historical representation of multiwavelength observations for the source SDSSJ1539+3954. Row 1.) X-ray observations from Chandra ACIS. Row 2.) ZTF Band-g light curve. A maximum variability of only about $\sim$0.3 $\pm$ 0.03 magnitude is observed for the quasar from the ZTF light curve, a level of variability that is typical among quasars. Row 3.) Maroon data points: VLA-FIRST observation at 1.4 GHz, Red data points: GMRT observations at 1.26 GHz, Yellow data points: VLASS observations at 3 GHz, Indigo data point: VLA observation at 3 GHz. The Chandra epochs are marked with vertical lines (orange) that show concurrent variability in X-ray and optical wavebands. The GMRT Band-5 observation epochs are marked with dot-dashed vertical lines (cyan).}
    \label{fig:allobs}
\end{figure*}

Furthermore, Figure \ref{fig:allobs} shows the light curves of the source at X-ray, optical, and radio wavelengths. The top panel shows the archival X-ray observations from \citet{ni2022sensitive}. The soft-band X-ray flux exhibited a significant brightening in the 2019 observation as compared to its undetected state in the 2013 observation. Shortly thereafter, the source dimmed again, returning to its X-ray weak, undetected state. 
The ZTF g-band light curve (middle panel) also shows a similar trend, with an initial brightening followed by a subsequent dimming around the same time period.  However, it should be noted that the ZTF magnitudes show a maximum variability of only about $\sim$0.3 $\pm$0.03 magnitude over the full period of ZTF observation for the source, a level of variability that is typical among quasars.  
Hence, the changes in optical magnitude do not show remarkable variability in their continuum emission (see Figure \ref{fig:allobs}). The third panel of \ref{fig:allobs} shows the 1.4 GHz and 3 GHz radio flux densities at various epochs, including measurements from archival surveys, VLA-FIRST and VLASS, as well as targeted observations using GMRT and VLA. The VLA-FIRST data (indicated with a brown square data point) obtained in the year 1994 show no detection of the source. The upper limit on the VLA-FIRST flux density is estimated using a 3$\sigma$ threshold on the rms background noise of the VLA-FIRST image. All the targeted observations conducted after 2020,  using GMRT and VLA, show clear detection of the source. However, the flux density at 1.4 GHz is 0.38 mJy, consistent with the VLA-FIRST 3$\sigma$ upper limit of 0.42 mJy. Although the 1.4 GHz measurements from VLA-FIRST and GMRT are consistent, we cannot rule out variability since the VLA-FIRST value is an upper limit. Similarly, the source is not detected in any of the VLASS survey images due to faint radio emission from the source remaining below the background noise level. The VLA Band-S (3 GHz) flux density obtained in 2022 (represented by the indigo star data point in \ref{fig:allobs}) is low enough to remain consistent with the VLASS non-detection.  

Finally, our spectroscopic follow-up data confirm no changes in the source’s weak emission-line nature.  \citet{ni2022sensitive} examined optical spectra from SDSS and the Hobby--Eberly Telescope (HET) observations, reporting no significant changes in emission-line strength from 2004 to 2020. Consistent with these findings, our HCT follow-up observations in 2020 and 2021 also reveal that all emission-lines remain weak. Figure \ref{fig:opticalspectra} (in Appendix \ref{app:specs}) presents the observed optical spectra from the HCT, spanning the rest wavelength range of 1400 $\mathring{\mathrm{A}}$ to 2700 $\mathring{\mathrm{A}}$. While the emission-lines persist as weak, some continuum dimming is evident, aligning with the trends seen in the ZTF light curves.
\section{Discussion}\label{sec:discussion}
In this study, we primarily investigated the radio properties of the radio-quiet weak emission-line quasar SDSSJ1539+3954, with an emphasis on exploring any potential relationship between its remarkable X-ray variability and its radio emission. Using data from both archival surveys (VLA-FIRST and VLASS) and targeted observations (GMRT and VLA), we measured the radio flux densities and derived the spectral index. While no significant variability in radio flux was identified, the targeted observations confirmed faint radio emission consistent with earlier non-detections. In this section, we discuss the implications of our analysis.

\subsection{Correlated X-ray and radio variability? }
Several studies (e.g., \citealt{2008ApJ...689...79C, Chatterjee_2009, 2011MNRAS.411..402B, 2011ApJ...734...43C,2022MNRAS.510..718P,10.1093/mnras/stac1891}) have explored the disk--jet connection in AGNs by examining correlations between X-ray and radio variability, often finding the radio lagging the X-rays by weeks to years. The variability amplitudes in both bands are frequently comparable, suggesting a physical link between X-ray emission and jet activity. However, exceptions exist--such as NGC 4051 \citep{10.1093/mnras/stw2810}--where strong X-ray variability is not accompanied by corresponding radio changes. Young radio sources with recently triggered jets often experience rapid variations in their radio flux over the time scale of a few years as the jets interact with the surrounding interstellar medium \citep{1996ApJ...473L..13F, 2016ApJ...818..105M, 2020ApJ...897..128K, Wolowska_2021}. Thus, if the X-ray and radio emission from the quasar SDSSJ1539+3954 are indeed connected, substantial radio variability would be expected on multi-year timescales, particularly if the variability was caused by the ejection of new material.

Prior to the first X-ray variability event in 2019, SDSSJ1539+3954 is observed at 1.4 GHz in the VLA-FIRST survey (epoch 1994) and at 3 GHz in the VLASS survey (epoch 2017), with no detections in either case. The multi-epoch VLASS 3 GHz observations, spanning 2017, 2020, and 2023, likewise present a consistent picture over six years, with no detections above the VLASS sensitivity limit ($\approx$120 $\mu$Jy/beam RMS per epoch, or $\approx$360 $\mu$Jy/beam at 3$\sigma$). GMRT multi-band observations from 2020, 2022, and 2024 show no statistically significant variability that could be attributed to delayed variations in the radio emission. The 3 GHz flux density in 2020 can therefore be assumed to be similar to that measured from the VLA observation in 2022 ($\approx$159 $\mu$Jy). However, the clear detections in GMRT and VLA observations raise the possibility that the source may have undergone radio variability relative to the VLA-FIRST epoch.

A critical window for constraining variability is the period between 2019 and 2020, during which the source exhibits a dramatic X-ray dimming by a factor of $\sim$9. Importantly, this interval is bracketed by VLASS 3 GHz observations in 2017 (prior to the dimming), 2020 (during the decay), and 2023, providing multi-epoch radio coverage that can be used to assess any correlated variability. If the radio and X-ray emissions are correlated, such a change would plausibly be accompanied by a decrease in radio flux. Even under a moderate correlation, the radio flux during the X-ray--bright epoch in 2019 would be expected to exceed the VLASS detection threshold. For instance, a ninefold increase would imply a flux above 1.4 mJy--easily detectable in the 2017 VLASS epoch. However, the source is not detected in VLASS 2017, despite the rest-frame time difference between 2017 and 2019 epochs being only $\sim$234 days ($\sim$7.7 months). Even for a smaller variability amplitude, such as a factor of 2.5, the expected 2017 flux ($\sim$398 $\mu$Jy) remains above the VLASS detection threshold. The absence of a detection in 2017 thus places meaningful constraints on the amplitude of radio variability during the X-ray-bright phase.

We note one caveat that the 2017 VLASS epoch precedes the observed peak X-ray brightness in 2019 by only $\sim$7.7 months in the rest frame. However, high-amplitude radio variability over such short timescales is not commonly observed in AGN, particularly at GHz frequencies. Additionally, the ZTF optical light curve remains stable between 2017 and 2019, with no evidence of significant optical variability that might indicate major changes in accretion or jet activity during that time. These observations argue against any significant rise and fall in radio emission that would correlate with the 2019 X-ray flare, and support our conclusion that no correlated radio variability is observed over the relevant epochs.

Another important consideration is the large spatial scales probed by low-frequency radio observations, particularly in the GMRT bands. At these frequencies, the synthesized beam encompasses both compact nuclear regions and any extended diffuse structures within the host galaxy or its surroundings. When such diffuse emission spans the entire region sampled by the beam, it can dominate the integrated flux, potentially obscuring variability originating from the compact core. This blending effect makes it inherently difficult to isolate and detect changes in the nuclear component, especially when its contribution to the total flux is relatively small. However, even allowing for contamination from diffuse emission, a ninefold increase in the compact component in 2017--relative to its measured value in 2020--should have produced a significant and measurable enhancement in the total integrated flux. The absence of such a signature in the 2017 VLASS data argues strongly against a substantial radio response correlated with the X-ray flare.

Overall, we conclude that the radio emission from SDSSJ1539+3954 remains relatively stable over timescales of several years and does not exhibit variability comparable to that observed in the X-rays during the period of strong X-ray changes. Therefore, any direct connection between the radio and X-ray emission in this source is not evident. However, we cannot rule out the possibility that the source may have exhibited a small-scale radio variability over a longer timescale of approximately 26 years, between the VLA-FIRST epoch and the GMRT epoch. 
The time difference between the X-ray flux rise and our latest radio observations corresponds to a rest-frame timescale of 1.72 years, which translates to a physical scale difference of approximately 0.5 pc between the X-ray and radio-emitting regions. This distance corresponds to the region downstream of the jet where it becomes optically thin. It would be valuable to continue monitoring this source in the future. Thus, an alternative, and perhaps more likely, scenario involves the presence of separate X-ray-emitting and radio-emitting regions at different scales within the AGN. This would imply that the observed X-ray variability and radio emission from the quasar are not correlated.

\subsection{Similarity with other WLQs}
\citet{Wang_2024} recently reported extreme X-ray variability in the WLQ SDSSJ152156.48+520238.4 ($z\sim 2.24$) between 2006 and 2023. The quasar’s soft X-ray (0.5--2 keV) flux varied by a factor of $\approx$ 32 within 0.97 rest-frame years, peaking in 2023 Chandra observations. Infrared and optical observations during this period showed only mild variability. The source, monitored with Chandra and XMM-Newton in 2006, 2013, 2019, and 2023, exhibited two notable rises in X-ray flux: an initial increase through 2013, followed by a decline in 2019, and a subsequent brightening to its highest state in 2023 (see Figure 3 and Table 2 in \citealt{Wang_2024}).
Furthermore, \citet{10.1111/j.1365-2966.2012.21648.x} and \citet{Liu_2022} have also reported similar dramatic X-ray variability observed in WLQs PHL 1092 ($z \sim$0.4) and  SDSSJ1350+2618 ($z \sim 2.6$), respectively. The quasar PHL 1092, in its XMM-Newton observations from the year 2000 to 2010, exhibited a dramatic drop in its 2 keV X-ray flux by a factor of $\sim$260 from 2003 to 2008 followed by a rise in its flux by a factor of $\sim$26 in 2010  (see Table 1 and Figure 1 in \citet{10.1111/j.1365-2966.2012.21648.x}). The WLQ SDSSJ1350+2618 displayed a rise in its soft X-ray flux (0.5 - 2 keV) by a factor of approximately 6, followed by a dimming by a factor of about 7.6, as observed in its Chandra observations from 2015 to 2016 (see Table 1 and Figure 1 in \citet{Liu_2022}).  

To investigate the radio-emission properties of the WLQs, as mentioned earlier, we examined available radio data from various radio surveys, as well as archival targeted radio observations of the sources. For the quasar SDSSJ152156.48+520238.4, we analyzed data from the VLA-FIRST and VLASS surveys, as well as archival targeted VLA observations. The VLA-FIRST (at 1.4 GHz, 1993 to 2011) and VLASS (at 3 GHz, 2017 to 2023) surveys provide quick-look flux density images, while the VLA archival data includes observations at 6 GHz (C-band) in 2022 and 10 GHz (X-band) in 2023. As with SDSSJ1539+3954, the source is undetected in both the VLA-FIRST and VLASS survey images. Likewise, the targeted VLA observations at 6 and 10 GHz frequencies also show no detection of the source. Together with the findings for SDSSJ1539+3954, these results suggest no clear connection between extreme X-ray variability and radio emission in WLQs.

Conversely, the other two WLQ sources that exhibited X-ray variability--namely, PHL 1092 and SDSSJ1350+2618 -- do not have any available radio data from targeted observations following the X-ray variations. However, we note that both sources are classified as radio-quiet, with only PHL 1092 detected in the FIRST survey at 1.4 GHz. Thus, it is difficult to examine their radio properties in the context of the X-ray variability events. 

\subsection{X-ray variability and Thick-Disk plus Outflow (TDO) model}\label{subsec:tdo}
\citet{ni2018connecting} proposed a thick-disk plus outflow (TDO) model to explain the observed X-ray variability in WLQs like SDSSJ1539+3954. This model suggests that super-Eddington accretion systems may develop a geometrically thick inner accretion disk, resulting in high column densities and disk outflows that obscure high-energy ionizing photons (X-ray/EUV) from reaching the high-ionization broad line region (BLR) of the AGN; see Figure 1 of \citep{ni2018connecting}. According to the TDO model, the inner accretion disk’s thickness varies azimuthally, creating geometric asymmetry. When the observer’s line of sight passes through the thicker regions of the disk, the quasar appears X-ray weak, while it appears X-ray normal otherwise. Consequently, the extreme X-ray variability observed in SDSSJ1539+3954 can be attributed to structural changes in the accretion disk. Our finding of no correlation between X-ray and radio variability aligns with the TDO model’s explanation, as the model attributes X-ray variability primarily to geometric and structural factors within the accretion disk rather than an intrinsic luminosity change or to a direct coupling with radio emission.

\subsection{Origin of radio emission in WLQ SDSSJ1539+3954}
As discussed in \citet{Panessa2019} (and references therein), the non-thermal radio emission from RQ-AGNs can arise from various mechanisms, including (1) high star formation rates; (2) particle acceleration in the AGN corona; (3) AGN outflows/winds interacting with the surrounding medium; and (4) small-scale jets. The optically thick radio emission from the core of the AGN or AGN corona exhibits a flat radio spectrum ($\alpha \geq -0.5$) and follows the observed correlations between GHz radio emission and X-ray emission in the 0.2 to 20 keV range \citep{Brikmann2000, giroletti, Arilaor}. Conversely, the radio emission resulting from synchrotron radiation in extended outflows or jets displays a steep spectrum ($\alpha \leq -0.5$), with a spectral break indicating the aging of the accelerated charged particles \citep{peterson1997introduction, netzer2013physics}. 

In cases where radio emission is dominated by starburst activity, one observes low-frequency ($\nu < 30$ GHz) synchrotron emission with steep GHz spectra (e.g., $\alpha \approx -0.7$) that transition to flatter spectra ($\alpha \approx -0.1$) at higher frequencies ($\nu > 30$ GHz), producing diffuse, unresolved emission on galactic scales \citep{1992ARA&A..30..575C, Panessa2019}.
Nevertheless, the brightness temperature ($T_B$) of the total radio emission in unresolved/compact radio sources can be used to differentiate between radio emission originating from star formation or AGN activity. For star formation, $T_B$ is typically limited up to $\sim 10^5$ K \citep{1992ARA&A..30..575C, 10.1093/mnras/stab549, 10.1093/mnras/stac2129, 10.1093/mnras/staf020}. Similarly, outflowing AGN winds interacting with the interstellar medium (ISM) also exhibit diffuse, low-brightness temperature radio emission \citep{2006AJ....132..546G, Panessa2019}. 

We computed the brightness temperature using the method outlined in \citet{2005ApJ...621..123U}:
\begin{equation}
    T_B = 1.8 \times 10^9 (1+z) \left( \frac{S_\nu}{1 \, \text{mJy}} \right) \left( \frac{\nu}{1 \, \text{GHz}} \right)^{-2} \left( \frac{\theta_1 \theta_2}{1 \, \text{mas}^2} \right)^{-1} K
\end{equation}

Since the source remains unresolved in all the frequency bands, we have used the obtained clean beam sizes as the upper limit on $\theta_1$ and $\theta_2$ (see sections \ref{subsec:GMRT_analysis} and \ref{subsec:vla_analysis}). Thus, for the source parameters, $z\simeq 1.935$ and the obtained flux densities at several frequencies (see Table \ref{tab:table-2020-GMRT}), the lower limit on estimated $T_B$ ranges from $\sim$1230 K to 28 K across the entire frequency range (340 MHz to 10 GHz). Notably, the source is not resolved in any radio observation, with the maximum resolution obtained from the radio observations being $\sim$0.33 arcsec. Thus, we have obtained lower limits on the obtained brightness temperature estimates in all the radio frequency bands.   

Our radio follow-up observations of WLQ SDSSJ1539+3954 reveal an unresolved source with a steep spectrum and no detectable short-term radio variability. Below, we discuss the possible mechanisms driving its radio emission based on these observed characteristics.

\textbf{\underline{A small-scale jet}:} 
The radio spectral index of the quasar shifts from approximately $\alpha \sim -0.65$ at lower frequencies to $\alpha \sim -1.09$ above 3GHz, corresponding to an approximate steepening rate of $0.87\beta = -0.27$ per GHz near 1GHz (see Figure \ref{fig:radio-plots} and Section \ref{sec:results} for details). This spectral steepening indicates optically thin synchrotron emission from aging electrons, consistent with small-scale jetted radio emission. However, at a resolution of 0.23" and $z\sim1.935$, the projected scale of the radio emission is $\lesssim$2 kpc,\footnote{\url{https://www.astro.ucla.edu/~wright/CosmoCalc.html}} placing an upper limit on the size of the extended region. While the current resolution does not allow a definitive conclusion regarding the jet's presence, the observed low brightness temperatures rule out the young jet scenario. Radio emission from jets or outflowing blobs typically exhibits a brightness temperature greater than $> 10^8$ K at GHz frequencies \citep{Giroletti_2009,2013ApJ...765...69D,panessajet,10.1093/mnras/stab587,10.1093/mnras/stab2651}. Nevertheless, VLBA observations (which we expect to acquire in the coming months) of the source with a higher mas resolution in GHz bands can uncover the nature and extent of radio emission in the source. The long-baseline VLBA observations at 1.4 and 6 GHz provide high angular resolutions of approximately 5 mas and 1.6 mas, respectively, for the quasar SDSSJ1539+3954. At these resolutions, we can probe source rest-frame spatial scales of up to $\sim$40 pc and $\sim$20 pc in the respective bands, allowing us to distinguish between radio emission originating from AGN-core-jet structure as well as star formation/AGN outflows--driven radio emission.

\textbf{\underline{Star formation/Outflowing AGN wind}:} The source SDSSJ1539+3954 remains unresolved in both VLA and GMRT observations, meaning the measured radio flux densities may include contributions from the AGN core as well as extended structures in the host galaxy, such as AGN winds or star-forming regions. The quasar exhibits a steep power-law spectral index ($\sim$-0.65 before spectral break and $\sim$-1.09 after the spectral break) and low brightness temperature limits, both consistent with radio emission arising from either star formation or AGN winds, specifically at GHz radio frequencies. Our observations cannot rule out either possibility, while the higher resolution VLBA observations can reveal the true nature of the radio emission.  

However, considering that WLQ SDSSJ1539+3954 is a high-accretion source and our radio observations support the TDO model, the AGN wind is likely the dominant mechanism driving the radio emission in the quasar. 

\textbf{\underline{AGN-core/Coronal radio emission}:} 
As mentioned previously, radio emission from the AGN core or corona typically produces a flat spectrum, whereas the quasar exhibits a steep spectral index, ruling out core or coronal emission as the primary mechanism. Additionally, if the radio emission were coronal in origin, we would expect a significant correlation between radio and X-ray fluxes since both are produced in the same region; however, no such correlation is observed. The steep spectrum, lack of radio variability, and absence of a radio--X-ray correlation lead us to rule out coronal radio emission as the primary source. However, it is interesting to note that the source satisfies the condition $L_R/L_X \approx 10^{-5}$ \citep{Arilaor} for coronal radio emission in RQ-quasars. The 0.2-20 keV X-ray luminosity ($L_X = 4.09 \times 10^{45}$ ergs/s) and 5 GHz radio luminosity ($L_R = 1.67 \times 10^{41}$ ergs/s) are obtained from the Chandra observations (X-ray normal epoch) and the VLA observations, respectively. Furthermore, considering the light travel time scales and absence of variability in our three GMRT observations, spanning a period of 4 years, constrains the emitting region to scales of at least 0.42 pc -- significantly larger than the expected size of a corona for a $10^8 M_{\odot}$ black-hole (even if the size of the corona is assumed to be extended to 100 $r_g$, where $r_g$ is the gravitational radius). Thus, the radio-emitting region extends to farther scales than the corona. Hence, the X-ray and the dominant mechanism driving radio emission from the quasar are not coupled.

\section{Conclusions}\label{sec:conclusion}

In this study, we carried out follow-up radio observations of SDSSJ1539+3954, a radio-quiet weak-line quasar known for its extreme X-ray variability from 2013 to 2020. While the source was undetected in the VLA-FIRST survey, targeted observations with the GMRT (2020, 2022 and 2024) and VLA (2022) telescopes revealed a point-like radio source. No significant variability is observed across multiple GMRT epochs, and the source remains undetected in all three VLASS (2017, 2020 and 2023) epochs.
The overall radio dataset, characterized by non-detections in all VLASS epochs and consistent flux densities from GMRT and VLA observations, all within the 3$\sigma$ sensitivity limits of FIRST and VLASS, supports the interpretation that the radio emission has remained largely stable over time. However, a radio variability over a scale of $\sim 26$ years between VLA-FIRST and GMRT observations can't be ruled out. 
This suggests that the X-ray and radio emissions are driven by separate processes, consistent with the thick-disk plus outflow model proposed for weak-line quasars.
The steep spectral index excludes dominant radio emission from the corona. The observations are consistent with the radio emission originating from either an outflowing AGN wind or the star formation phase of the host. Higher-resolution observations, such as those achievable with the VLBA, would help further clarify the nature of the quasar’s radio emission.

\section*{Acknowledgment}
MV acknowledges support from the Department of Science and Technology, India - Science and Engineering Research Board (DST-SERB) in the form of a core research grant (CRG/2022/007884). WNB acknowledges support from the Chandra X-ray Center grant G02-23083X. AJN would like to acknowledge DST-INSPIRE Faculty Fellowship (IFA20-PH-259) for supporting this research. PK acknowledges the support of the Department of Atomic Energy, Government of India, under the project 12-RandD-TFR-5.02-0700. The authors thank the operation teams of GMRT-NCRA-TIFR and VLA-NRAO facilities, who made the radio observations possible. GMRT is run by the National Center for Radio Astrophysics of the Tata Institute of Fundamental Research. VLA is run by the National Radio Astronomy Observatory. The National Radio Astronomy Observatory is a facility of the U.S. National Science Foundation operated under cooperative agreement by Associated Universities, Inc.
AC acknowledges support from Rupesh Behera, and Prerna Biswas from IIA, India, and Salmoli Ghosh from NCRA-TIFR, India, for their help in the data analysis.

\newpage
\appendix 
\section{Optical Spectra} \label{app:specs}
\vspace{-0.5cm}
\begin{figure}[h!]
    \centering
    \hspace{-0.4cm}
    \includegraphics[scale=0.45]{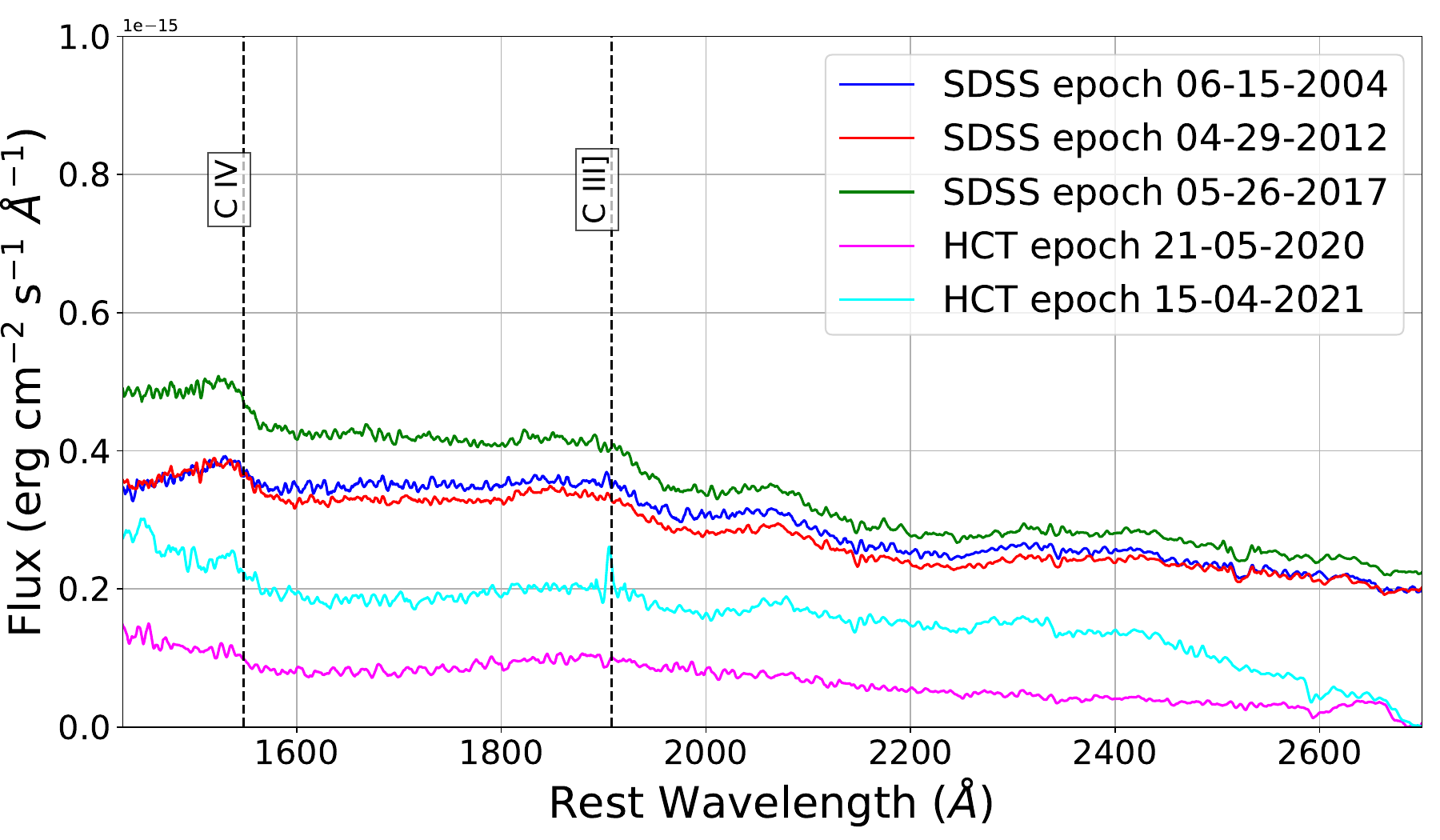}
    \caption{SDSS and HCT spectra of the WLQ SDSSJ1539+3954 in the rest wavelength range 1400 $\mathring{\mathrm{A}}$ to 2700 $\mathring{\mathrm{A}}$. Spectra obtained from the SDSS-DR16 catalog include observations of the source in the years 2002 (blue), 2012 (red), and 2017 (green). The spectra obtained in follow-up observations in 2020 and 2021 using HCT HFOSC are shown in magenta and cyan colors, respectively. No change can be observed in the spectrum and the emission-line signatures of the source.}
    \label{fig:opticalspectra}
\end{figure}
\bibliography{References}
\bibliographystyle{aasjournalv7}

\end{document}